\documentclass[11pt]{article}
\usepackage{moriond,epsfig}

\def\be{\begin{equation}}
\def\ee{\end{equation}}
\def\bea{\begin{eqnarray}}
\def\eea{\end{eqnarray}}

\newcommand{\mh}{\mbox{$m_{\rm{h}}$}}
\newcommand{\mH}{\mbox{$m_{\rm{H}}$}}
\newcommand{\mw}{\mbox{$m_{\rm{W}}$}}
\newcommand{\mt}{\mbox{$m_{\rm{t}}$}}
\newcommand{\mz}{\mbox{$m_{\rm{Z}}$}}
\newcommand{\mA}{\mbox{$m_{\rm{A}}$}}
\newcommand{\mhch}{\mbox{$m_{\rm{H}^\pm}$}}
\newcommand{\Ecm}{\mbox{$E_{\rm{cm}}$}}
\begin{document}
\vspace*{4cm}
\title{EXPERIMENTAL SUMMARY}

\author{ GAIL G. HANSON }

\address{Department of Physics, Indiana University,\\
Bloomington, Indiana U.S.A.}

\maketitle\abstracts{The experimental results presented at the XXXVI
Rencontres de Moriond Electroweak Interactions and Unified Theories
are summarized. The results range from possible evidence for the
Standard Model Higgs boson at LEP2 to searches for new physics at LEP,
the Tevatron and HERA, to precision electroweak  and weak decay 
measurements, and to the
strong evidence for neutrino oscillations.}

\section{Introduction}

This XXXVIth Rencontres de Moriond Electroweak Interactions and
Unified Theories offered some intense speculations on the recent past
and the next few years, occurring just a few months after the shutdown
of LEP and at the beginning of the new Tevatron run at Fermilab.
There were 49 experimental contributions~\cite{gen} to this conference,
so I cannot possibly do justice to them all. I can only try to touch
on what seemed to me to be some of the highlights.

This summary begins with the tantalizing evidence from LEP for the
Standard Model Higgs boson, and then continues with the precision
electroweak and top quark mass measurements from LEP, the Tevatron,
SLD, BEPC, and HERA. Searches for non-Standard-Model Higgses at LEP and
prospects for Higgs searches at future colliders are then
discussed. Searches for supersymmetry and other new physics at LEP,
Fermilab, and DESY are reviewed. The measurement of positron
polarization at PSI is summarized, and the AMANDA and DAMA searches
for extraterrestrial interactions are briefly mentioned, as are the
antimatter experiments at CERN. The beautiful evidence for neutrino
oscillations from Super-Kamiokande and preliminary K2K results are
presented. Rare kaon and charm decays are summarized. The LEP and SLD
$B$ physics measurements are then discussed. The measurements of the $CP$
violation parameter $\sin 2 \beta$ from Belle and BaBar are presented,
followed by measurements of rare $B$ decays from CLEO, Belle, and
BaBar. Finally the precision measurement of the anomalous magnetic
moment of the muon from Brookhaven National Laboratory E821 is summarized.

\section{Evidence for the Standard Model Higgs Boson}

The results of searches for the Standard Model Higgs boson at LEP were
presented by Nakamura,~\cite{OPAL_higgs}
Morettini,~\cite{DELPHI_higgs} and
Teixeira-Dias.~\cite{LEP_higgs} 
At LEP the SM Higgs boson is expected to be produced mainly through
the Higgs-strahlung process $e^+ e^- \rightarrow H^0 Z^0$, with
contributions from the $WW$ fusion channel below 10\%. Searches are
performed in the  channels
$HZ \rightarrow b \bar b q \bar q$ (four jet),
$HZ \rightarrow b \bar b \nu \bar \nu$ (missing energy),
$HZ \rightarrow b \bar b \tau^+ \tau^-$ or $\tau^+ \tau^- q \bar q$ (tau),
and
$HZ \rightarrow b \bar b e^+ e^-$ or $b \bar b \mu^+ \mu^-$
(leptonic). All four LEP experiments have published results of the
searches using the 2000 data. ALEPH, DELPHI, and OPAL presented
results here that are the same as their publications and very similar
to those presented at the November 3, 2000, meeting of the LEP
Experiments Committee (LEPC), for which the only official combination of the
four LEP experiments exists. The combination was presented by
Teixeira-Dias.~\cite{LEP_higgs} A summary of the comparison of the
presentations here with those used in the combination is given in 
Table~\ref{higgs}.

\begin{table}[htb]
\caption{Comparison of results presented here (published) with
those used in the November 3 LEP combination. \label{higgs}}
\vspace{0.4cm}
\begin{center}
\begin{tabular}{|l|c|c|} \hline
{\bf Experiment}               & {\bf Nov. 3 LEPC} & {\bf
               Presented here (published)}     \\
 & {\bf (combination)} & \\ \hline
ALEPH      & $3.4 \sigma$    & $3.2 \sigma$ ($1 - CL_b$ = 0.0015 at
               \mH\ $\sim$ 115 GeV)                   \\ \hline
DELPHI     & $-1.0 \sigma$    & $-1.0 \sigma$, \mH\ $> 114.3$~GeV
               (113.5~GeV expected)           \\ \hline
L3       & $1.7 \sigma$    & -- (reanalysis by summer conferences)   \\ \hline
OPAL  & $1.3 \sigma$    & $1.3 \sigma$, \mH\ $> 109.7$~GeV (112.5 GeV
               expected) \\
 & & signal slightly favored          \\ \hline
\end{tabular}
\end{center}
\end{table}

\begin{figure}[htb!]
\begin{center}
\centerline{\epsfig{file=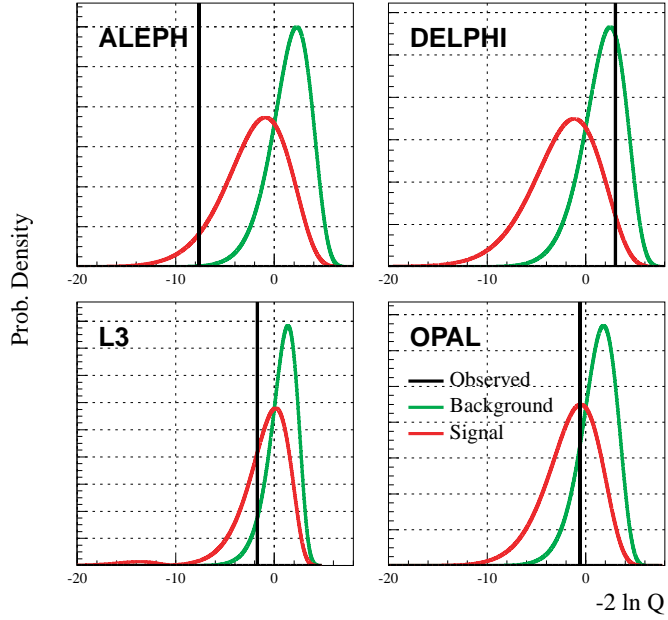,width=0.55\textwidth}}
\caption{Probability density distributions for each LEP experiment for
signal and background for \mH\ = 115~GeV as shown at the November 3
LEPC meeting. One experiment is very
signal-like, two experiments are slightly signal-like, and one
experiment is background-like.
\label{probsb}}
\end{center}
\end{figure}

\begin{figure}[htb!]
\begin{center}
\centerline{\epsfig{file=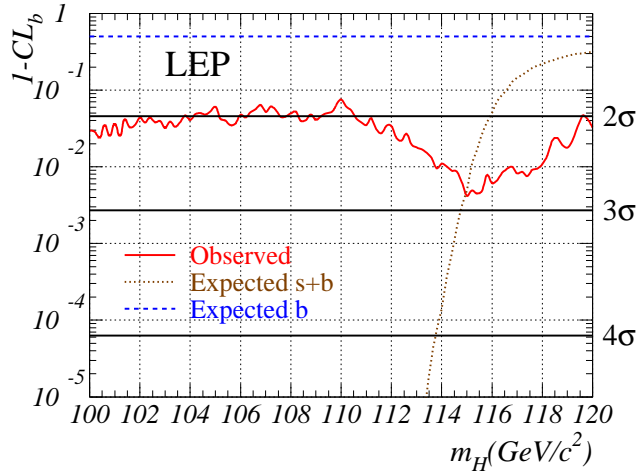,width=0.55\textwidth}}
\caption{$1-CL_b$ for the combined LEP experiments
as shown at the November 3 LEPC meeting.
\label{clb}}
\end{center}
\end{figure}

\begin{figure}[htb!]
\begin{center}
\centerline{\epsfig{file=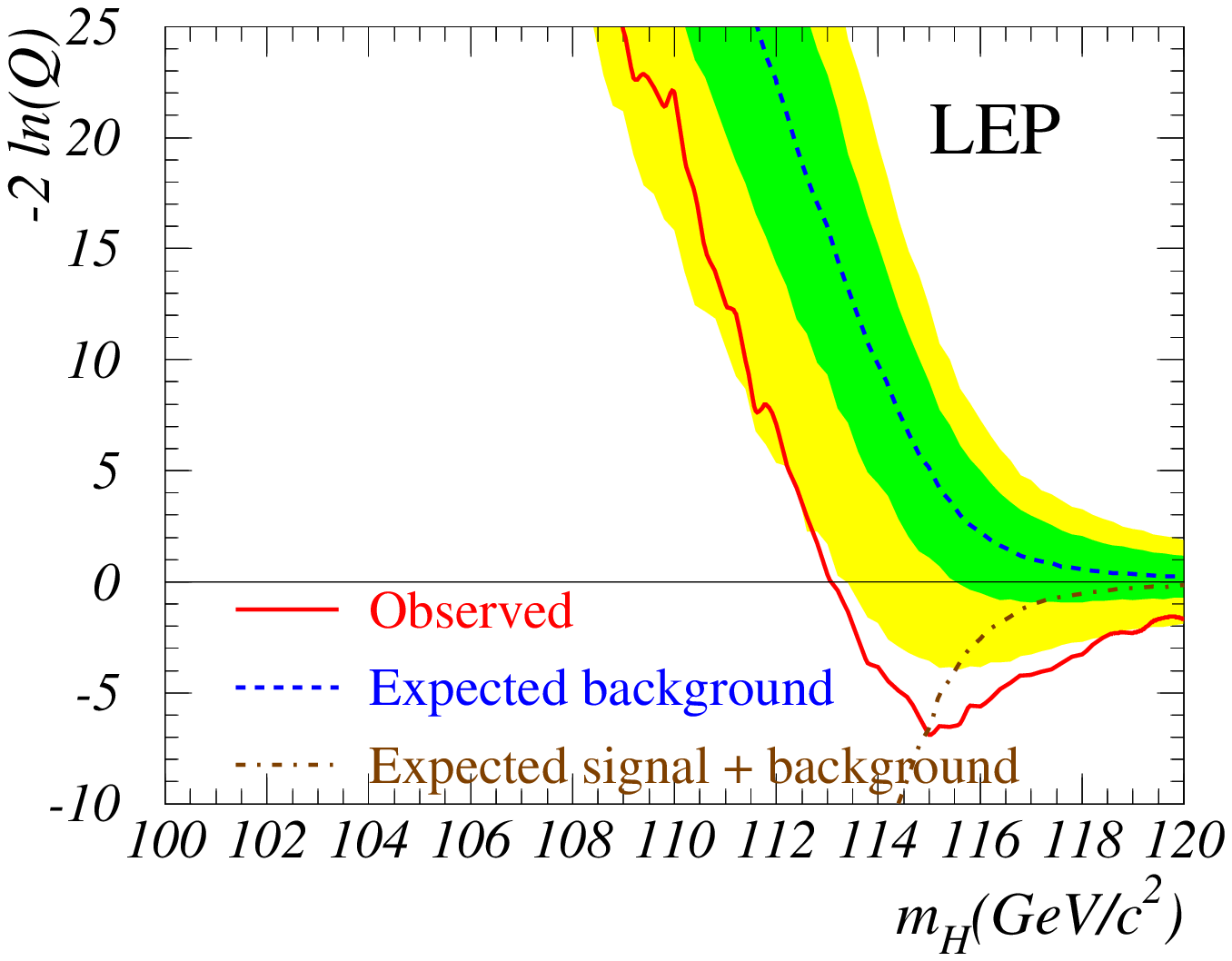,width=0.55\textwidth}}
\caption{Log likelihood distributions for the combined LEP experiments
as shown at the November 3 LEPC meeting. 
\label{lnq}}
\end{center}
\end{figure}

The combination of the four LEP experiments presented at the November
3 LEPC meeting showed an excess of $2.9 \sigma$ significance, or $1 -
CL_b$ = 0.0042 at \mH\ $\sim$ 115 GeV. The value of the mass is \mH\ =
115 $^{+1.3}_{-0.9}$ GeV ($2\sigma$ errors). The probability density
distributions for signal and background for \mH\ = 115 GeV for each
experiment as presented at the November 3 LEP meeting are shown in 
Fig.~\ref{probsb}. The confidence level for background for the LEP
combination is shown in
Fig.~\ref{clb}. For background only, $1-CL_b$ will be 0.50 on the average.
The log likelihood distribution for the LEP combination is shown in
Fig.~\ref{lnq}. The value of \mH\ is given by the point at which the
observed log likelihood intersects the predicted signal plus
background curve.

 The final combination of the four LEP experiments will be produced
after the L3 Experiment reanalyzes its data and releases the results,
in time for the 2001 summer conferences.

\section{Electroweak and Top Measurements at LEP, SLD and Tevatron}

The masses of the $W$ boson and the top quark obtained in Tevatron
Run~I by CDF and D0 were presented by Glenzinski:~\cite{doug} the
combined values are \mw\ =
80.452 $\pm$ 0.062 GeV and \mt\ = 174.3 $\pm$ 5.1 GeV. Watson
presented \mw\ = 80.446 $\pm$ 0.040 GeV, a preliminary combination
based on 82\% of all LEP2 data. The world average, including LEP1,
SLD, $\nu N$, and \mt\ is \mw\ = 80.368 $\pm$ 0.023
GeV. Verzi~\cite{verzi} presented results on $W W$, $Z Z$, $Z \gamma$
and single $W$ production (triple gauge boson couplings) from LEP. 
Quartic gauge boson coupling results were presented by Musy.~\cite{musy}

\begin{figure}[ht!]
\begin{center}
\centerline{\epsfig{file=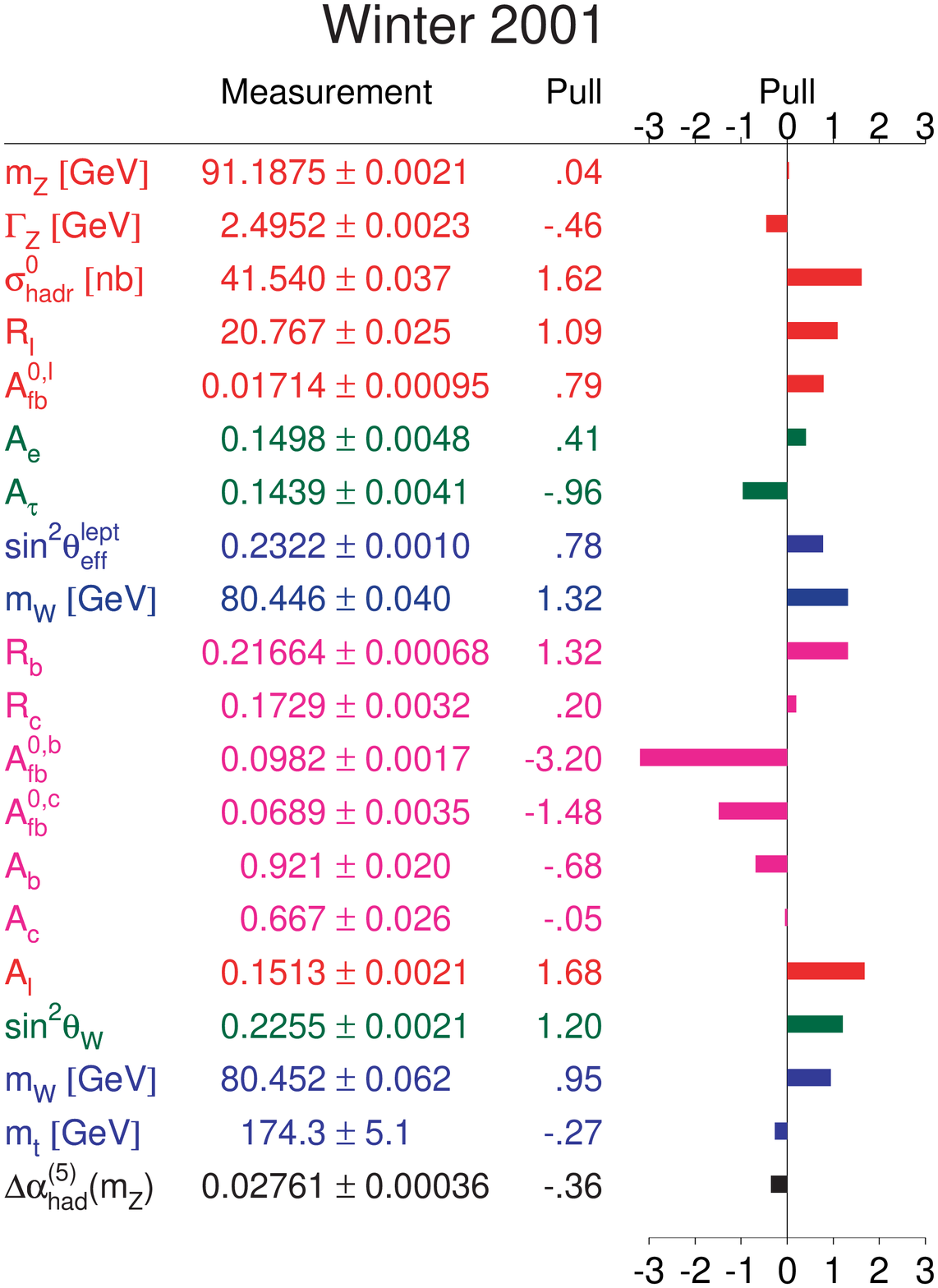,width=0.55\textwidth}}
\caption{Fit to precision electroweak data pulls.
\label{pulls}}
\end{center}
\end{figure}

De Groot~\cite{degroot} presented new or updated results on  $R_b$, 
$R_c$, $A_b$, and $A_c$ from SLD. The new SLD average for  $A_b$ is
0.913 $\pm$ 0.021, compared with the LEP average of 0.873 $\pm$ 0.018,
derived from the measurements of the $b \bar b$
forward-backward asymmetry ($A_{\rm FB}^{0,b}$) and $A_e$.

\begin{figure}[!htb]
\begin{center}
\centerline{\epsfig{file=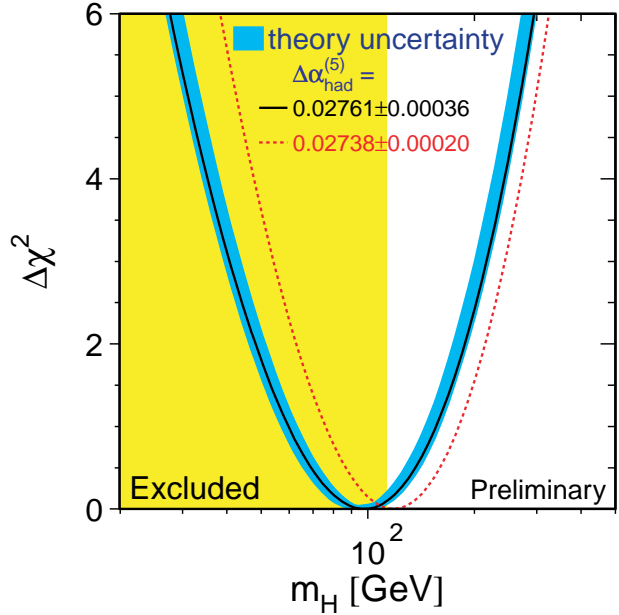,width=0.55\textwidth}}
\caption{Mass of Standard Model Higgs boson $vs.$ $\chi^2$ of fit.
\label{chsqmh}}
\end{center}
\end{figure}

Tournefier~\cite{lepew} reported on the fits of the LEP electroweak
data to the Standard Model including
the running of $\alpha_s$, the latest \mw\ from LEP2, the new $R$
measurement from BES, the new measurements of $A_{\rm FB}^{0,b}$
from ALEPH and DELPHI, and the new values
of $R_b$, $R_c$, $A_b$, and $A_c$ from SLD. 
The results of the fit are
shown in Fig.~\ref{pulls}. The $\chi^2$ of the fit, including the top
mass measurement, as a function of \mH\ is shown in Fig.~\ref{chsqmh}.
The fit gives \mH\ = 98$^{+58}_{-38}$ GeV and \mH\ $<$ 212 GeV at 95\%
C.L. However, the probability of the fit is only 4\%
($\chi^2$/d.o.f. = 25/15). The largest deviation is due to $A_{\rm
FB}^{0,b}$. The data favor a light Higgs boson.

\clearpage

\section{{\it R} Measurement at BES}

\begin{figure}[!hbt]
\begin{center}
\centerline{\epsfig{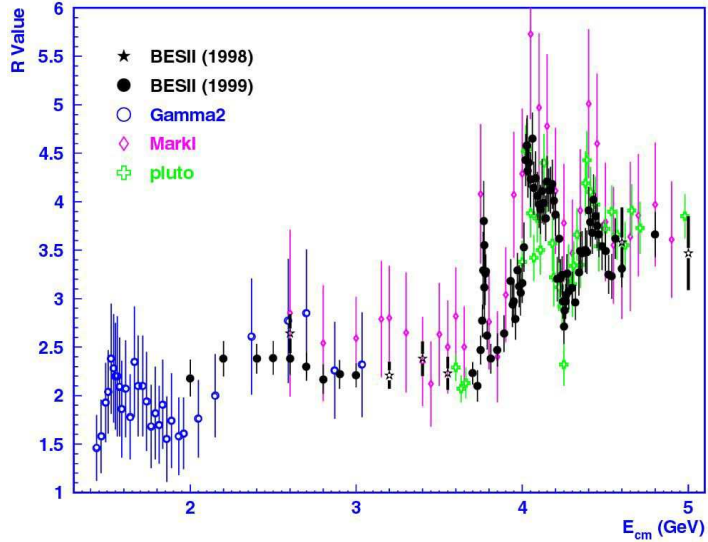}}
\caption{$R$ values in the \Ecm\ range 1-5 GeV.
\label{rbes}}
\end{center}
\end{figure}

A precision measurement of
\[R = \frac{\sigma(e^+ e^- \rightarrow hadrons)}{\sigma(e^+ e^-
\rightarrow \mu^+ \mu^-)} \]
from 2 to 5 GeV center-of-mass energy (\Ecm) reduces the uncertainty on
$\alpha (m^2_Z)$, reduces the width of $\chi^2$ $vs.$ \mH\ in the
electroweak fits, and reduces the uncertainty in the theoretical value
of the muon ($g-2$). Previous measurements of $R$ had uncertainties of 15-20\%
in the 2-5 GeV \Ecm\ range. A new set of measurements was made using
the Upgraded BEijing Spectrometer (BES-II) at the Beijing
Electron-Positron Collider (BEPC) and presented by Huang.~\cite{bes} 
The uncertainties in the new
measurements range from 6 to 10\% (average 6.6\%), a factor of two
improvement over the old measurements. Values of $R$ {\it vs.} \Ecm\
are shown in Fig.~\ref{rbes}.

\section{Electroweak Tests at High $Q^2$ at HERA}

Fusayasu~\cite{ewhera} presented recent results on high-$Q^2$ neutral
and charged currents at HERA. Measurements of $xF_3^{\rm NC}$ have
been extracted from the difference of the neutral current differential
cross sections for $e^- p$ and $e^+ p$. The first space-like measurement of
\mw\ has been extracted from the slopes of the charged current
differential cross sections for $e^\pm p$. Electroweak
unification is observed from the equality of the neutral current and
charged current cross sections at high $Q^2 \gg$  \mw$^2$, \mz$^2$.
One fb$^{-1}$ integrated luminosity is
expected from a luminosity upgrade with left-handed and right-handed
longitudinally polarized beams in $2001-2005+$.

\section{Non-SM Higgses at LEP}

Teuscher~\cite{teuscher} and Holzner~\cite{holzner} discussed the
status of searches for non-Standard Model Higgs bosons at LEP. For the
fermiophobic Higgs boson ($h^0 \rightarrow \gamma \gamma$) the
preliminary LEP combined limit is \mh\ $>$ 107.7 GeV, 95\% C.L. L3 has
also combined their search for $h \rightarrow \gamma \gamma$ with a
search for $h \rightarrow W W^*$ to obtain a limit of \mh\ $>$ 104.8
GeV.

Preliminary combined MSSM Higgs search results give 0.52 $<$
tan$\beta$ $<$ 2.25 excluded and limits on the $h^0$ and $A^0$ masses
of \mh\ $>$ 89.9 GeV and \mA\ $>$ 90.5 GeV in the
conservative \mh-max scenario.

In general Type II Two-Higgs Doublet Models (2HDM(II)), Higgs boson
decays to $b \bar b$ and $\tau^+ \tau^-$ can be suppressed, leaving $c
\bar c$ and $gg$ decays dominant. Flavor-independent searches for
Higgs bosons have been carried out and give preliminary limits
of \mh\ $>$ $102-109$ GeV assuming Standard Model cross sections for
Higgs boson production. Complete flavor-independent 2HDM(II) scans
over the parameter space of the Higgs boson masses \mh\ and \mA, 
tan$\beta$ and $\alpha$ have been performed. For any value of
$\alpha$ between $-\pi/2$ and 0, the region 13 $<$ \mA\ $<$ 56 GeV and
\mh\ $<$ 44 GeV is excluded for  0.4 $\leq$ tan$\beta$ $\leq$ 58.

Searches for Higgs bosons decaying into invisible particles (e.g., $h
\rightarrow \tilde{\chi}^0 \tilde{\chi}^0$) give \mh\ ~$>$ 113.7~GeV for
a Standard Model Higgs boson production cross section and 100\%
branching ratio for decay into invisible particles.

In the MSSM \mhch\ $>$ 330 GeV from the absence of flavor-changing neutral
currents (from $b \rightarrow s \gamma$ measurements). Thus searches
for $H^\pm$ at LEP2 are non-MSSM searches. For tan$\beta$ $<$ 1,
decays to $c \bar s$ and $\tau^+ \nu$ dominate, whereas for
tan$\beta$ $>$ 1, $W^* A$ dominates. L3 has observed for years
a significant excess
(now $4.4 \sigma$) at 68 GeV in the four-jet channel, but the other LEP
experiments have not confirmed it. The combined preliminary LEP limit
for $H^\pm \rightarrow c \bar s$ and $\tau^+ \tau^-$ is
\mhch\ ~$>$~78.5 GeV. 
A search for $H^\pm \rightarrow W^* A$ has been performed by
OPAL with no significant excess observed.  

\section{Prospects for Higgs at Future Colliders}

Upgrades to the luminosity of the Tevatron at Fermilab were discussed
by Church.~\cite{Tev_lumi} Run~IIa started the week before the
conference with the goal of 2 fb$^{-1}$ by 2003. Run IIb will begin in
2003 with the luminosity goal of 13 fb$^{-1}$ by 2006$+$, and many
upgrades (mostly increasing antiproton production) will be implemented
to achieve this goal:

\begin{itemize}
\item Radiation due to beam losses will be decreased so that the
Booster intensity can be increased.
\item Slip stacking will be used to increase the intensity in the Main
Injector.
\item The $\bar p$ yield will be increased by increasing the gradient
of the lithium lens.
\item $\bar p$'s are cooled in the Debuncher Ring, the Accumulator
Ring and the Recycler. Higher bandwidth stochastic cooling will be
used in the Debuncher and the Accumulator. Electron cooling in the
Recycler Ring will be installed in 2003.
\item Electron lens compensation (an electron beam coaxial with the $\bar p$
beam) will be used to compensate the
beam-beam tune shift that limits the proton intensity in the
Tevatron. The first electron lens is installed in one straight section
and will be commissioned over the next few months.
\item The bunch spacing will be reduced from 396 ns to 132 ns, which
will also require introducing a crossing angle at the IR's.
\item The Tevatron beam energy will be increased to 980 GeV. 
\end{itemize}
\noindent
The total luminosity goal by 2006 with all upgrades is 15  fb$^{-1}$.

In Run II at the Tevatron~\cite{higgs_fnal} the Standard Model 
Higgs boson will be
searched for in associated production in the modes $\ell \nu b \bar
b$, $e^+ e^- b \bar b$, and $\nu \bar \nu b \bar b$ for 90 $<$ \mH\ $<$
130 GeV. For 130 $<$ \mH\ $<$ 190 GeV, the Higgs decay modes searched
will be $WW^*$ and $ZZ^*$. For an integrated luminosity of  15
fb$^{-1}$, the Higgs boson signal should be significant by $5 \sigma$
for \mH\ $<$ 120 GeV. A $3 \sigma$ effect can be observed for \mH\ $<$
130 GeV and 150 $<$ \mH\ $<$ 175 GeV, and a 95\% C.L. upper limit can
be determined for \mH\ $<$ 185 GeV.

Costanzo~\cite{higgs_lhc} presented the prospects for Higgs searches
at the LHC. The SM Higgs can be searched for in the channels $H
\rightarrow \gamma \gamma$ and $t \bar t H \rightarrow t \bar t b \bar
b$ for \mH\ $<$ 130 GeV, in $H \rightarrow ZZ^* \rightarrow 4 \ell$ for
\mH\ $>$ 130 GeV, and in $H \rightarrow WW$ and $WH \rightarrow WWW$
for \mH\ $>$ 160 GeV. The Higgs boson can be discovered at $5 \sigma$
for ATLAS plus CMS together for \mH\ $<$ 130 GeV with 10  fb$^{-1}$ 
integrated luminosity per experiment ($\sim$ one year, the first year
of physics running, 2006-2007). 
For 100 fb$^{-1}$ (four years running) the Higgs signal will be
significant at $10 \sigma$ per experiment over the entire \mH\ mass 
range.

Schumacher~\cite{lc_higgs} discussed Higgs physics at a future $e^+
e^-$ linear collider, which would have a center-of-mass energy of
350-800 GeV and a luminosity of 2-5 $\times 10^{34}$ cm$^{-2}$
s$^{-1}$ (several 100 fb$^{-1}$/year). An  $e^+ e^-$ linear 
collider would be complementary to the LHC:  the LHC would discover an
SM Higgs boson for 100 $<$ \mH\ $<$ 1000 GeV and could discover at least one
MSSM Higgs boson, and the linear collider would measure the properties,
such as mass, width, and couplings, with more precision than the
LHC. For example, \mH\ could be measured to $\sim$ 50~MeV for \mH\
$\sim$ 150 GeV. Branching ratios to $b \bar b$, $gg$, $WW$, etc., can
be measured to $\sim$ 5\%. $CP$ properties can be measured from
angular distributions in $e^+ e^- \rightarrow ZH$. Holes in the LHC
region for discovery of MSSM Higgs bosons can be closed.

Murray~\cite{mc_higgs} presented the possibilities with a muon
collider Higgs factory. The beam energy in a muon collider can be
measured to 10$^{-6}$ via ($g-2$), and a  10$^{-5}$ beam energy spread 
is possible. Since muons are 200 times heavier than electrons, the
storage ring can be smaller. The Higgs boson is produced through the
$s$-channel, so the cross section is 40,000 times the electron
equivalent. With such a machine one could measure \mH\ to 100's of keV
and the width to 1 MeV. $CP$ properties can be measured from
transverse polarization asymmetries. A muon collider Higgs factory can
offer many advantages in the case of MSSM or other non-standard-model
Higgs bosons through, for example, precise measurement of the Higgs
width and resolution of degenerate states. However, several technical
difficulties must be overcome. The muons decay, so cooling of the
beams must take place quickly. The proposed method is ionization
cooling. Ionization cooling schemes typically cool in the transverse
phase space. This cooling then needs to be transferred to the
longitudinal direction by emittance exchange. These cooling ideas are
being developed but have not yet been shown to be feasible.

\section{Searches for SUSY and Exotics at LEP}

Jakobs~\cite{jakobs} presented preliminary results on MSSM SUSY
searches, including data from 2000. With gravity mediated SUSY
breaking and R-parity conservation, the lightest supersymmetric
particle (LSP) is assumed to be the lightest neutralino
($\tilde{\chi}^0_1$). The combined LEP limits for slepton searches are
$m_{\tilde{e}} > 99.4$ GeV, $m_{\tilde{\mu}} > 96.4$ GeV, and 
$m_{\tilde{\tau}} >
87.1$ GeV, all for $m_{\tilde{\chi}^0_1} = 40$~GeV. There was no
$\tilde{\tau}$ excess in the 2000 data. The combined LEP limits for
squarks are $m_{\tilde{t}_1} >$ 95 GeV for $\tilde{t}_1 \rightarrow c 
\tilde{\chi}^0_1$ and $\Delta M = 40$ GeV, $m_{\tilde{t}_1} >$ 97 GeV 
for $\tilde{t}_1 \rightarrow b \ell \tilde{\nu}$ and $\Delta M = 40$
GeV, and $m_{\tilde{b}} >$ 95 GeV 
for $\tilde{b} \rightarrow b \tilde{\chi}^0_1$ and $\Delta M = 20$
GeV, all independent of mixing angle. The limit on the mass of the
lightest chargino is $m_{\tilde{\chi}^+_1} > 103.5$ GeV for $\Delta M
>$ 10 GeV. The low $\Delta M$ regions are covered by dedicated searches.
The limit on the mass of the LSP is $m_{\tilde{\chi}^0_1} > 39.6$ GeV
for large $m_0$. A new DELPHI result includes slepton and Higgs
searches.

De Min~\cite{demin} summarized the preliminary results of searches 
at LEP for R-parity violating (RpV) SUSY, gauge mediated SUSY breaking, 
and exotic final states. In gauge mediated SUSY breaking, $\tilde{\chi}^0_1
\rightarrow \tilde{G} \gamma$ or $\tilde{\ell} \rightarrow \tilde{G}
\ell$, where $\tilde{G}$ is the gravitino. Single $\gamma$ final
states are often used. Searches have also been performed for light
sgoldstinos ($\Phi$): $e^+ e^- \rightarrow \Phi \gamma$, with $\Phi
\rightarrow gg$ or $\gamma \gamma$. Unfortunately, no significant
excesses have been seen. A new LEP Working Group on Exotic Searches
has been formed.

\section{New Phenomena at the Tevatron}

\begin{figure}[htb!]
\begin{center}
\centerline{\epsfig{file=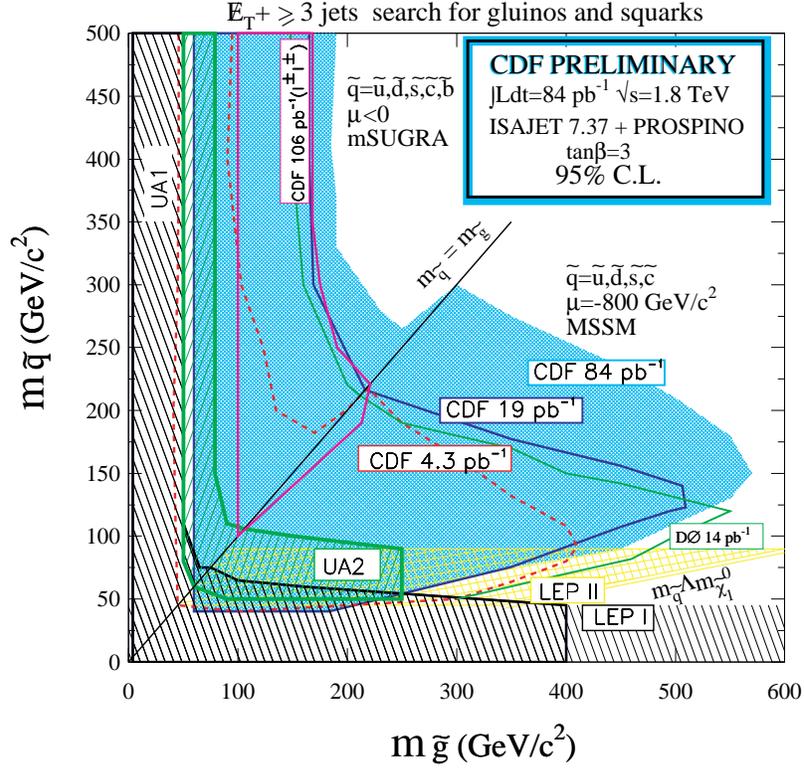,width=0.70\textwidth}}
\caption{New CDF search for squarks and gluinos.
\label{tevsusy}}
\end{center}
\end{figure}

CDF and D0 collected $\sim$ 110 pb$^{-1}$ integrated luminosity during the
Tevatron Run I $1992-1996$. B\"{u}scher~\cite{buscher} reported the
results of a new MSSM scalar top search by D0 in the mode $\tilde{t}_1
\rightarrow b \ell \tilde{\nu}$. The excluded region will be greatly
extended with Run II data. CDF searched for RpV scalar top in the mode
$\tilde{t}_1 \rightarrow \tau^+ b$. D0 carried out a search for large
extra dimensions using di-fermion/di-photon events and found
consistency with Standard Model expectations; limits were set at the
$\geq$ 1 TeV level. D0 has developed a model independent search
(SLEUTH) for
excesses at high-$p_T$ and has tested it for Monte Carlo samples of
new physics, such as leptoquarks, and on data. Berryhill~\cite{berryhill} 
reported on a model independent search developed by CDF based on
leptons, photons, multiple jets and missing transverse energy ($E_T$). A new
CDF search for scalar quarks excludes $m_{\tilde{q}} \sim m_{\tilde{g}} <
300$ GeV, as shown in Fig.~\ref{tevsusy}. A search for events with 
electrons, large missing $E_T$, and large mass of the 
electron-missing-$E_T$ system excludes $m_{W^\prime} < 755$
GeV and compositeness scales $\Lambda_+$, $\Lambda_- < 2.8$ TeV. A D0
search for high-mass $ee$ events excludes $m_{Z^\prime} < 670$ GeV and
technicolor $m_{\rho_T}$,  $m_{\omega_T} < 207$ GeV.

\section{Searches for Extra Dimensions at LEP}

Searches for extra dimensions were also carried out at LEP, as
presented by Gataullin.~\cite{gataullin} Direct searches consisted of
looking for one $\gamma$ or one $Z^0$ plus missing energy due to a
missing graviton ($G$), and indirect searches looked for deviations from
(d$\sigma$/d$\Omega$)$_{\rm SM}$ due to $G$ exchange ($\lambda$ = $\pm
1$ refers to the different signs of the interference term). All four
LEP experiments searched for $e^+ e^- \rightarrow G^*
\rightarrow \gamma \gamma$ with combined limits on the fundamental
mass scale $M_S > 0.95$ TeV
for $\lambda = +1$ and  $M_S > 1.14$ TeV for $\lambda = -1$. OPAL and
L3 searched for deviations in  $e^+ e^- \rightarrow e^+ e^-$ giving
combined limits of  $M_S > 1.13$ TeV
for $\lambda = +1$ and  $M_S > 1.28$ TeV for $\lambda = -1$. OPAL
combined $ee$, $\gamma \gamma$ and $ZZ$ searches to obtain $M_S > 1.03$ TeV
for $\lambda = +1$ and  $M_S > 1.17$ TeV for $\lambda = -1$.

\section{Searches Beyond the Standard Model at HERA}

\begin{figure}[htb!]
\begin{center}
\centerline{\epsfig{file=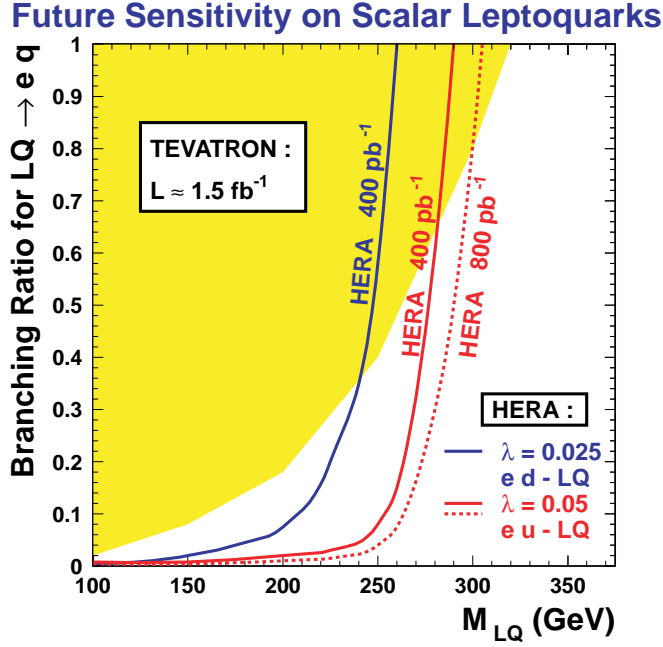,width=0.55\textwidth}}
\caption{Future prospects for leptoquark searches at HERA.
\label{futurehera}}
\end{center}
\end{figure}

New search results from HERA were presented by
Sch\"{o}ning~\cite{schoning} for leptoquarks, large extra dimensions,
lepton flavor violation, R-parity violating supersymmetry, and excited
fermions. H1 has observed an excess of events with isolated leptons
with large transverse momentum and large missing transverse momentum
in both the $1994-97$ $e^+p$ data and the $1999-2000$ $e^+p$
data. ZEUS, however, finds consistency with Standard Model expectations.
Leptoquarks are excluded up to $\sim$ 260 GeV. New R-parity violating
limits in mSUGRA improve the limits from the Tevatron for Yukawa
couplings $\lambda_{1j1} \sim \alpha_{em}$. New limits on anomalous
production of top quarks were obtained: HERA is very sensitive to such
production. The HERA upgrade starts in September and will produce five
times higher luminosity and polarized $e^\pm$ beams. In four years
running, an integrated luminosity of 800 pb$^{-1}$ is anticipated,
giving a large potential to discover leptoquarks, as shown in 
Fig.~\ref{futurehera}.

\section{NuTeV: Heavy Lepton Anomaly}

Shaevitz reported results from the NuTeV experiment at
Fermilab.~\cite{nutev} The experiment uses 800 GeV protons on target
to produce $\nu_{\mu}$ and $\bar \nu_{\mu}$ beams and measures sin$^2$
$\theta_W$ = 0.2253 $\pm$ 0.0019 $\pm$ 0.0010 (preliminary),
corresponding to M$_W$ = 80.26 $\pm$ 0.110 GeV. They have implemented
a decay detector consisting of a 35 m helium bag instrumented with drift
chambers to search for heavy neutral leptons. They observed three $\mu
\mu$ events with vertices in the helium and transverse mass $>$ 2.2
GeV compared with an expected
background of 0.04 events. They do not observe any $\mu e$ or $\mu
\pi$ events although the expected background for these is about three
times larger. The events have the characteristics of neutrino
interactions because the energies are asymmetric. They do not seem to
be consistent with any known background or other physics
process. The events are used to set limits for heavy neutral lepton and 
neutralino production.

\section{Polarization of $e$ from $\mu$ Decays at PSI}

The purposes  of the $\mu$P$_T$ Experiment at PSI, discussed by
K\"{o}hler,~\cite{epol} are a search for additional couplings
in muon decays, a model independent determination of $G_F$, and a
search for violation of time-reversal invariance. All three components
of the positron's polarization vector are measured. No evidence for
additional scalar couplings in muon decay were found. The model
independent measurement of $G_F$ gives 
\[G_F = (1.16532 \pm 0.00069) \times 10^{-5} (\hbar c)^3 {\rm GeV}^{-2}. \]

\section{AMANDA, DAMA and Antimatter Experiments}

Hill~\cite{amanda} reported on the latest results from AMANDA,
which is an experiment at the South Pole that looks at atmospheric
neutrinos that penetrate the Earth. The results are consistent with
expectations.  They have also set limits on extraterrestrial
neutrinos, WIMPs, and gamma-ray bursts.

DAMA is an experiment that uses NaI(Tl) to search for dark matter
sources. Belli~\cite{dama} gave the current status of the analysis,
which shows an annual modulation with the proper features to be
compatible with the presence of WIMPs in the Galactic halo. No
systematics or side reactions were found to be able to give the
observed modulation.

Tan described several antimatter experiments at 
CERN.~\cite{antimatter} Low energy antiprotons (5.3 MeV) are extracted from the
Antiproton Decelerator (AD) Ring. The three experiments are ATRAP,
ATHENA, and ASACUSA. ATRAP has produced the world's coldest $\bar
p$'s.  The $\bar p$'s are trapped in the
world's most intricate Penning Trap. ATRAP also accumulated the first
cold positrons. The cold $\bar p$'s and cold $e^+$'s are then used to
produce cold antihydrogen for high resolution spectroscopy. 
Experiments studying transitions in antihydrogen as compared
with hydrogen test $CPT$ invariance.

\section{Neutrinos}

\subsection{$\nu$ Mass Measurement at Mainz}

The best measurement of the $\nu_e$ mass is being carried out at Mainz
using the tritium $\beta$ decay endpoint. Bonn~\cite{mainz} reported
that they have the
world's best sensitivity and have measured $m_{\nu_e}$ $<$ 
2.2~eV/$c^2$, 95\% C.L., using 1998/99 data. They have seen no evidence of 
the ``Troitsk anomaly.'' 
Changing the design of their apparatus will allow them to reach
a sensitivity of $\le$ 0.4 eV/$c^2$.

\subsection{DONUT}

Sielaff~\cite{donut} reported results from the DONUT Experiment, 
Fermilab E872, which has observed four events due to
$\nu_{\tau}$ interactions in nuclear emulsion targets with an expected
background of 0.3 events.  The $\nu_{\tau}$ interaction
produces a $\tau$, which then decays to a single charged track and a
$\nu_{\tau}$ (86\% of $\tau$ decays). The decay results in a kink in
the reconstructed emulsion track. The probability that all four events
are background is 8.0 $\times$ 10$^{-5}$. Phase 2 of the analysis, with
relaxed cuts for the $\nu_{\tau}$ interaction, including three-prong
decays, is in progress. An interesting event was located on February
21, 2001.

\subsection{Super-Kamiokande Results}
\begin{figure}[!htb]
\begin{center}
\centerline{\epsfig{file=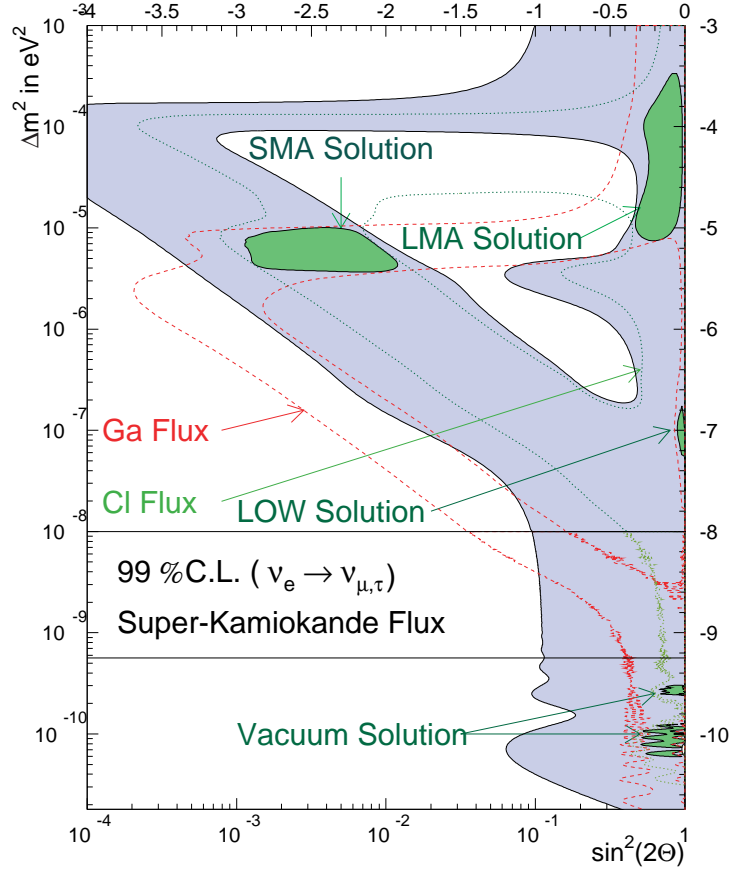,width=0.60\textwidth}}
\caption{Global fit to solar neutrino data from Super-Kamiokande for 
$\nu_e \leftrightarrow \nu_{\mu,\tau}$ oscillations.
\label{sk_solar}}
\end{center}
\end{figure}

Super-Kamiokande results on solar neutrinos were presented by
Smy.~\cite{sk_solar} Super-Kamiokande has reached its design threshold
of 5 MeV. The total observed  flux compared with the Standard Solar Model is
0.451$\pm$0.005 (stat.)$^{+0.016}_{-0.014}$ (syst.). The day-night 
asymmetry is 1.5$\sigma$ away from zero. A new oscillation analysis 
has been performed, including
zenith angle and new
information on the $^8$B spectrum. The results of the global fit to
$\nu_e \leftrightarrow \nu_{\mu,\tau}$ are shown in
Fig.~\ref{sk_solar}. The Small Mixing Angle solution is
mostly disfavored at about 95\% C.L. Mixing to sterile neutrinos alone
is also disfavored at 95\% C.L.

Atmospheric neutrino results from Super-Kamiokande were presented by
Toshito.~\cite{sk_atmos} For 79~kton-years of data

\[{\rm Sub-GeV\ contained\ events:} \ \frac{(\mu/e)_{\rm
DATA}}{(\mu/e)_{\rm MC}} = 0.638 \pm 0.017 \pm 0.050 \] 

\[{\rm Multi-GeV\ events:} \ \frac{(\mu/e)_{\rm
DATA}}{(\mu/e)_{\rm MC}} = 0.675 ^{+0.034}_{-0.032} \pm 0.080. \] 

\noindent
A two-flavor oscillation analysis for $\nu_{\mu} \rightarrow \nu_{\tau}$ gives
$\sin^2 2\theta > 0.88$ and $1.6 \times 10^{-3} < \Delta m^2 < 4
\times 10^{-3}$~eV$^2$ at 90\% C.L. Pure $\nu_{\mu} \rightarrow \nu_s$
is disfavored at 99\% C.L. Three different analyses are consistent
with charged current $\nu_\tau$ appearance, with 74 events expected so
far.

\subsection{K2K Results}

The goal of K2K is to establish neutrino oscillations through
$\nu_\mu$ disappearance and $\nu_e$ appearance. The flight length is
250 km from KEK to Super-Kamiokande. The $\nu_\mu$ beam flux and
$\nu_e$ contamination are measured in the near detector at KEK. GPS is used
for time synchronization between the near and far
sites. Ishii~\cite{k2k} reported results for the data
from June 1999 to June 2000, corresponding to $2.29 \times 10^{19}$
pot. The number of fully contained  events observed in the fiducial
volume of Super-Kamiokande was 28, with $37.8^{+3.5}_{-3.8}$ events 
expected without oscillation. There is a deficit of $\sim$ 1 GeV
$\nu_\mu$ after the 250 km flight path at 90\% significance. The
experiment resumed in January 2001.

\subsection{Solar and Atmospheric Three $\nu$ Analysis}

A fit to the standard three neutrino mixing matrix, including CHOOZ
reactor data and the updated analyses of Super-Kamiokande data, was
presented by Montanino.~\cite{montanino} It was found that maximal
$\nu_\mu \leftrightarrow \nu_\tau$ mixing is preferred from
atmospheric $\nu$ data. Maximal $\nu_1 \leftrightarrow \nu_2$ mixing
is preferred from solar $\nu$ data. The best fit for solar plus
atmospheric plus CHOOZ data gives $U^2_{e3} = \sin^2\theta_{13} \cong
0$. There is a multiplicity of solutions to the solar $\nu$ problem.
 
\subsection{CHORUS and NOMAD Results}

CHORUS and NOMAD were designed to detect $\nu_\tau$ CC interactions:
$\nu_\tau + N \rightarrow \tau^- + X$. These short baseline experiments
at CERN have found no evidence for neutrino oscillations in the
cosmologically relevant region $\Delta m^2 > 1$ eV$^2$, as reported by
Cocco.~\cite{cocco} Mixing angles
down to $\sin^2(2\theta_{\mu \tau}) \approx O(10^{-4})$ have been explored.
The NOMAD analysis is completed, and CHORUS has started ``phase II''
emulsion data analysis.

\section{Rare Kaon Decays}

Recent results from NA48 were presented by
T. \c{C}uhadar-D\"{o}nszelmann.~\cite{cuhadar} Based on the 1999 $K_S$
high intensity data (2.3 $\times$ 10$^8$ $K_S$), BR($K_S \rightarrow
\gamma \gamma$) was measured to be ($2.58 \pm 0.36\ {\rm stat.}\ \pm 0.22\
{\rm syst.}$) $\times 10^{-6}$. Using the 1998 and 1999 data,
preliminary results are BR($K_S
\rightarrow \pi^+ \pi^- e^+ e^-$) =  ($4.3 \pm 0.2\ {\rm stat.}\ \pm 0.3\
{\rm syst.}$) $\times 10^{-5}$ with no asymmetry ($A^S_{\pi \pi e e}$
= ($-0.2 \pm 3.4\ {\rm stat.}\ \pm 1.4\ {\rm syst.}$)\%).  BR($K_L
\rightarrow \pi^+ \pi^- e^+ e^-$) =  ($3.1 \pm 0.1\ {\rm stat.}\ \pm 0.2\
{\rm syst.}$) $\times 10^{-7}$ with $A^L_{\pi \pi e e}$
= ($13.9 \pm 2.7\ {\rm stat.}\ \pm 2.0\ {\rm syst.}$)\%. Again using the
1999 $K_S$ high intensity data, no events survive after all cuts,
giving the final result BR($K_S \rightarrow \pi^0 e^+ e^-$) $<$ $1.4 
\times 10^{-7}$, 90\% C.L. 

Preliminary results using 300 $\times$ 10$^6$ events from the 1997
KTeV data 
were given by Bellantoni.~\cite{ktev} The charge asymmetry in $K_{e3}$
decays ($\delta_L$): $\delta_L$ =  ($3.320 \pm 0.058\ {\rm stat.}\ \pm 0.046\
{\rm syst.}$) $\times 10^{-3}$, giving a new world average of ($3.305
\pm 0.063$) $\times 10^{-3}$, in agreement with no $CP$ violation. The
charge radius for $K^0_L \rightarrow \pi^+ \pi^- e^+ e^-$ was measured
to be $<R^2>$ =  ($-0.047 \pm 0.008\ {\rm stat.}\ \pm 0.006\
{\rm syst.}$)~fm$^2$, which
implies a mass difference $m_s - m_d \sim$ 100 MeV. From one event
passing the cuts, a limit of BR($K^0_L \rightarrow \pi^0 \pi^0 e^+
e^-$) $<$ 5.4 $\times$ 10$^{-9}$, 90\% C.L., was found. BR($K^0_L
\rightarrow \mu^+ \mu^- \gamma$) = ($3.66 \pm 0.04\ {\rm stat.}\ \pm 0.07\
{\rm syst.}$) $\times$ 10$^{-7}$. This measurement puts the tightest
constraints on the $K^0_L \rightarrow \gamma^{(*)} \gamma^{(*)}$ 
vertex factor $\rho_{\rm CKM}$. The data are used to find limits on 
$\rho_{\rm CKM}$ for two models:
$\rho_{\rm CKM}$ $>\ -1.0$ using the BMS model and $>\ -0.2$ using the
DIP model. A limit was placed on the lepton flavor violating decay
$K^0_L \rightarrow \pi^0 \mu^\pm e^{\mp}$ at $<$ 4.40 $\times$ 10$^{-10}$
at 90\% C.L. for two events found in the data. The 1999 data will
improve the statistics by a factor of 2.5 for $\pi^0 e^+ e^-$,  $\pi^0
\mu^+ \mu^-$, and  $\pi^0 \mu^\pm e^{\mp}$ and by a factor of 3.2 for
$\pi^+ \pi^- e^+ e^-$, $\mu^+ \mu^- e^+ e^-$, and $e^+ e^- e^+
e^-$. However $K^0_L \rightarrow \pi^0 e^+ e^-$ and  $K^0_L
\rightarrow \pi^0 \mu^+ \mu^-$ are reaching background limits already.
Future goals are the measurements of $K^0_L \rightarrow \pi^0 \nu \bar
\nu$ in KaMI and $K^+ \rightarrow \pi^+ \nu \bar \nu$ in CKM.

\section{Charm Results}

New results for charm physics were presented by Link,~\cite{charm1}
Hans,~\cite{charm2} and Sanders.~\cite{charm3} The FOCUS 
Experiment~\cite{charm1} looks for
charm mixing through lifetime differences between $CP$-even and
$CP$-odd eigenstates in decays of the $D^0$. $D^0 \rightarrow K^+
\pi^-$ can occur through a doubly-Cabibbo-suppressed process or
through mixing followed by a Cabibbo favored decay. They measure a
lifetime difference $y$ = (3.42 $\pm$ 1.39 $\pm$ 0.74)\% and a ratio
of wrong-sign to right-sign branching ratios of (0.404 $\pm$ 0.085
$\pm$ 0.025)\% in the limit of no mixing; this is a possible
indication of $D^0 \bar D^0$ mixing at the $2 \sigma$ level. 

The CLEO Experiment~\cite{charm2} reported  $CP$ asymmetries consistent
with zero for five $D^0$ decays ($K^+ K^-$, $\pi^+ \pi^-$, $K^0_s
\pi^0$, $\pi^0 \pi^0$, and $K^0_s K^0_s$). They reported the first
observation of $D^0 \rightarrow K^- \pi^+ \pi^0$. They also measured
the time difference $y_{CP}$ = ($-0.011 \pm 0.025 \pm 0.014$) from
$K^+ K^-$ and $\pi^+ \pi^-$ decays as a limit on $D^0 \bar D^0$
mixing.

E791 at Fermilab~\cite{charm3} reported limits on many rare
or forbidden two-, three- and four-body decays of $D$ mesons, some of
which have not previously been reported. A few measurements are
improvements on the previous upper limits by an order of magnitude or more. 

\section {$B$ Physics from SLD and LEP}

Lin reported on recent heavy flavor results from LEP and
SLD.~\cite{SLD_LEP_B} A new preliminary measurement of the charm
multiplicity in $B$ decays from SLD gives $\eta_c$ = 1.238 $\pm$
0.055, which is in agreement with measurements from CLEO and LEP but
slightly higher and brings the average into better agreement with the
prediction. 

The measurements of the CKM matrix elements from LEP give
$|V_{cb}|$ = (40.4 $\pm$ 1.8) $\times$ 10$^{-3}$ (average of inclusive
and exclusive measurements) and $|V_{ub}|$ = (4.1 $\pm$ 0.7) $\times$
10$^{-3}$.
The $B^0_d$ mixing oscillation frequency is the most direct method to
extract $|V_{td}|$. The $B$ Oscillations Working Group (LEP, SLD and
CDF) value for $\Delta m_d$ is ($0.486 \pm 0.015$) ps$^{-1}$. Only a
lower limit has been obtained for $B^0_s$ mixing oscillations:
$\Delta m_s$ $>$ 15 ps$^{-1}$ at 95\% C.L., with a sensitivity of
18~ps$^{-1}$. These measurements can be used to constrain the
unitarity triangle, as discussed in the presentation by 
M. Ciuchini.~\cite{ciuchini} It will have to be left to Fermilab to
measure $\Delta m_s$.

\begin{figure}[htb!]
\begin{center}
\centerline{\epsfig{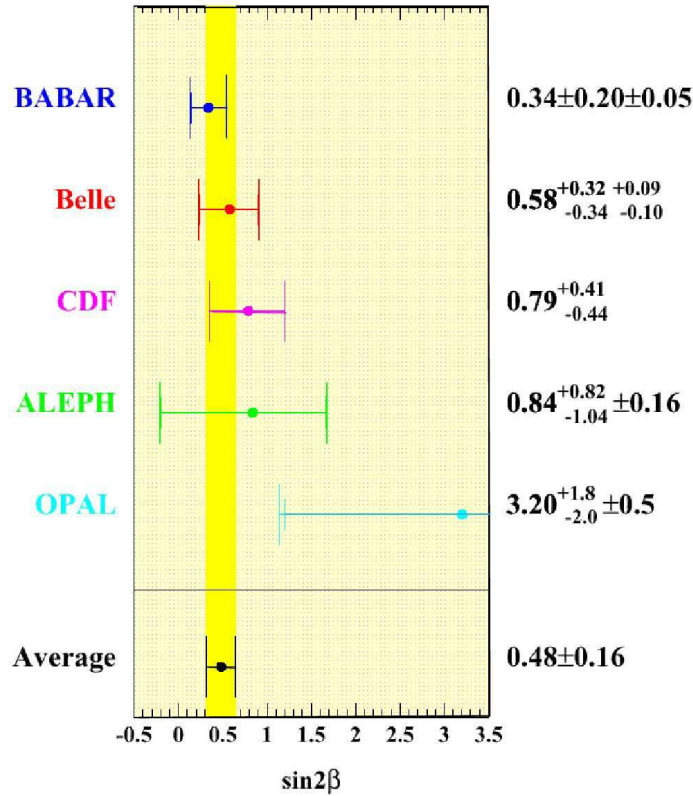}}
\caption{Compilation of measurements of the $CP$ violation parameter
$\sin 2\beta$.
\label{babar}}
\end{center}
\end{figure}
 
\section{$CP$ Violation and $B-\bar B$ Mixing Results from Belle and BaBar}

Belle, presented by Hanagaki,~\cite{hanagaki} and BaBar, presented by 
Faccini,~\cite{faccini} have made measurements
of the unitarity triangle angle $\sin 2\beta$ ($\phi_1 \equiv \beta$)
at the $B$ factories at KEK and SLAC, respectively:

\[{\rm Belle:} \ \sin 2\phi_1 = 0.58 ^{+0.32}_{-0.34}\ {\rm (stat.)}\
^{+0.09}_{-0.10}\ {\rm (syst.)} \] 

\[{\rm BaBar:} \ \sin 2\beta = 0.34 \pm 0.20\ {\rm (stat.)}\ \pm 0.05\ 
{\rm (syst.)} \]

Both experiments also measure $\Delta m_d$; the value for BaBar is 
$\Delta m_d =$ ($0.519 \pm 0.020$ (stat.) $\pm 0.016$ (syst.)) ps$^{-1}$
(preliminary) giving a new world average including Belle of 
$\Delta m_d =$ ($0.484 \pm 0.010$) ps$^{-1}$.

A compilation of $\sin 2\beta$ measurements is shown in Fig.~\ref{babar}.
After the next run ending August 2001 the uncertainty on  $\sin
2\beta$ should be 0.015.

\section{$B$ Decay Results from CLEO, Belle, and BaBar}

Results on rare $B$ decays were presented by CLEO, Belle, and
BaBar. CLEO has a new limit on $B^0 \rightarrow \pi^0 \pi^0$ and has
observed $B \rightarrow \phi K^{(*)}$, as presented by
Blanc.~\cite{blanc} The $b \rightarrow s \gamma$
branching ratio has been measured, and $|V_{cb}|$ has been determined
from the $\gamma$ energy spectrum to be  $|V_{cb}| =$ (40.6 $\pm 1.1$
(exp.) $\pm 0.7$ (theor.)) $\times 10^{-3}$ with a $CP$ asymmetry
$-0.27 < A_{CP} < +0.10$, 90\% C.L. Results for gluonic penguin,
electroweak penguin, and $B \rightarrow h h$ decays are shown in
Tables~\ref{gpeng}, \ref{ewpeng} and \ref{hh}, respectively.

Belle results were presented by Taylor,~\cite{taylor} based on 
an integrated luminosity of 10.5 fb$^{-1}$. They
have made the first observation of the Cabibbo-suppressed decays $B
\rightarrow D^+ K^-$, $D^{*0} K^-$, $D^{*+}K^-$, and $D^0
K^{*-}$. Measurements for $B \rightarrow \eta^{\prime} h$ give BR($B^+
\rightarrow \eta^{\prime} K^+$) = ($6.8 ^{+1.3 \ +0.7}_{-1.2 \ -0.9}$)
$\times 10^{-5}$ and BR($B^+
\rightarrow \eta^{\prime} \pi^+$) $ < 1.2 \times 10^{-5}$. Results for
electroweak penguin and $B \rightarrow h h$ decays are shown in
Tables~\ref{ewpeng} and \ref{hh}.

Roberts~\cite{roberts} reported BaBar results on $B$ decays to double charm.
BaBar took 20.7 fb$^{-1}$ of data on the $\Upsilon$(4S) resonance. The
branching ratio for $B^+ \rightarrow D^{*+} D^{*-} K^+$, a
color-suppressed decay, was measured to be ($0.34 \pm 0.16 \pm 0.09$)\%.
$B^0 \rightarrow D^{(*)+} D^{(*)-} K_S$, a potentially useful mode for
$CP$ violation measurements, was also observed. Charmless hadronic $B$
decays from BaBar were discussed by Wilson.~\cite{wilson} Results for 
gluonic penguin, electroweak penguin, and $B \rightarrow h h$ decays
are shown in Tables~\ref{gpeng}, \ref{ewpeng} and \ref{hh}.

All three experiments reported $CP$ asymmetry measurements that were
all consistent with zero. 

\begin{table}[htb]
\caption{Gluonic penguin decays. \label{gpeng}}
\vspace{0.4cm}
\begin{center}
\begin{tabular}{|l|c|c|} \hline
{\bf BR $\times 10^{-6}$}   & {\bf CLEO} & {\bf BaBar}     \\ \hline
                &                                &           \\
$\phi K^-$      & $5.5^{+2.1}_{-1.8} \pm 0.6$    & $7.7^{+1.6}_{-1.4}
\pm 0.8$  \\
                &                                &           \\
$\phi K^0$     & $< 12.3$, 90\% C.L.    &   $8.1^{+3.1}_{-2.5}
\pm 0.8$          \\
                &                                &           \\ 
$\phi K^{*-}$       &  $10.6^{+6.4 \ +1.8}_{-4.9 \ -1.6}$  &
$9.6^{+4.1}_{-3.3} \pm 1.7$ \\
                &                                &           \\ 
$\phi K^{*0}$       & $11.5^{+4.5 \ +1.8}_{-3.7 \ -1.7}$   & $-$ \\
                &                                &           \\
$\phi \pi^-$  & $-$ & $< 1.3$     \\ \hline
\end{tabular}
\end{center}
\end{table}

\begin{table}[htb]
\caption{Electroweak penguin decays. The Standard Model prediction is
(3.28 $\pm$ 0.33) $\times 10^{-4}$. \label{ewpeng}}
\vspace{0.4cm}
\begin{center}
\begin{tabular}{|l|l|c|} \hline
{\bf Experiment}   & {\bf Decay} &{\bf BR $\times 10^{-4}$} \\ \hline
CLEO      & $b \rightarrow s \gamma $    & $2.85 \pm 0.35 \pm 0.22$  \\
Belle     & $B \rightarrow X_s \gamma $  & $3.37 \pm 0.53 \pm 0.42
^{+0.50}_{-0.54}$ (theor.) \\
BaBar      &  $B^0 \rightarrow K^{*0} \gamma $ & $0.439 \pm 0.041 \pm 0.027$  \\ \hline
\end{tabular}
\end{center}
\end{table}

\begin{table}[htb]
\caption{Charmless hadronic $B$ decays. \label{hh}}
\vspace{0.4cm}
\begin{center}
\begin{tabular}{|l|c|c|c|} \hline
{\bf BR $\times 10^{-5}$}   & {\bf CLEO} & {\bf Belle} & {\bf BaBar}
\\ \hline
             &          &          &          \\
$K^+ \pi^-$      & $1.72^{+0.25}_{-0.24} \pm 0.12$ &
$1.87^{+0.33}_{-0.31} \pm 0.16$ & $1.67 \pm 0.16 ^{+0.12}_{-0.17}$
\\
             &          &          &          \\
$K^+ \pi^0$      &  $1.16^{+0.30 \ +0.14}_{-0.27 \ -0.13}$ &
$1.70^{+0.37 \ +0.20}_{-0.33 \ -0.22}$  & $-$  \\
             &          &          &          \\
$K^0 \pi^+$      & $1.82^{+0.46}_{-0.40} \pm 0.16$ &
$1.31^{+0.55}_{-0.46} \pm 0.26$  & $-$  \\
             &          &          &          \\
$K^0 \pi^0$      &$1.46^{+0.59 \ +0.24}_{-0.51 \ -0.33}$ &
$1.46^{+0.61}_{-0.51} \pm 0.27$ &  $-$  \\ 
             &          &          &          \\ \hline
             &          &          &          \\
$\pi^+ \pi^-$    & $0.43^{+0.16}_{-0.14} \pm 0.05$ &
$0.59^{+0.24}_{-0.21} \pm 0.05$ & $0.41 \pm 0.10 \pm 0.07$  \\
             &          &          &          \\ 
$\pi^+ \pi^0$    &  $< 1.27$, 90\% C.L. &
$0.71^{+0.36 \ +0.09}_{-0.30 \ -0.12}$ & $-$ \\
             &          &          &          \\ 
$\pi^0 \pi^0$     & $< 0.57$, 90\% C.L.    & $-$ & $-$ \\ \hline 
$K^+ K^-$       &  $< 0.19$, 90\% C.L.     & $-$ & $< 0.25$ \\ \hline
\end{tabular}
\end{center}
\end{table}

\section{Muon ($g-2$) Measurement at BNL E821}

The anomalous magnetic moment of the muon is given by $a_\mu
\equiv (g-2)/2$. For a point-like spin one-half particle, $g=2$.

\[ a_\mu^{\rm theory} = a_\mu^{\rm QED} + a_\mu^{\rm had} + a_\mu^{\rm
weak} + a_\mu^{\rm New\ Physics?} \]

\noindent
where 

\[ a_\mu^{\rm SM} =  a_\mu^{\rm QED} + a_\mu^{\rm had} + a_\mu^{\rm
weak}. \]

\noindent
The contributions~\cite{czarn} to $a_\mu^{\rm SM}$ are shown in 
Table~\ref{muanom}.

\begin{table}[htb]
\caption{Contributions to the muon anomalous magnetic moment. \label{muanom}}
\vspace{0.4cm}
\begin{center}
\begin{tabular}{|l|c|c|} \hline
{\bf Contribution} & {\bf $\times 10^{-11}$} & {\bf ppm}     \\ \hline
                   &                         &                     \\
$a_\mu^{\rm QED}$      & $116584706 \pm 3$    & $10^6 \pm 0.025$ \\
$a_\mu^{\rm had}$     & $ 6739 \pm 67 $    &   $57.8 \pm 0.6$     \\ 
$a_\mu^{\rm weak}$     &  $151 \pm 4 $  & $1.30 \pm 0.03$      \\ \hline
$a_\mu^{\rm SM}$       & $116591596 \pm 67$   & $\pm 0.6$ \\   \hline
\end{tabular}
\end{center}
\end{table}

\begin{figure}[!htb]
\begin{center}
\centerline{\epsfig{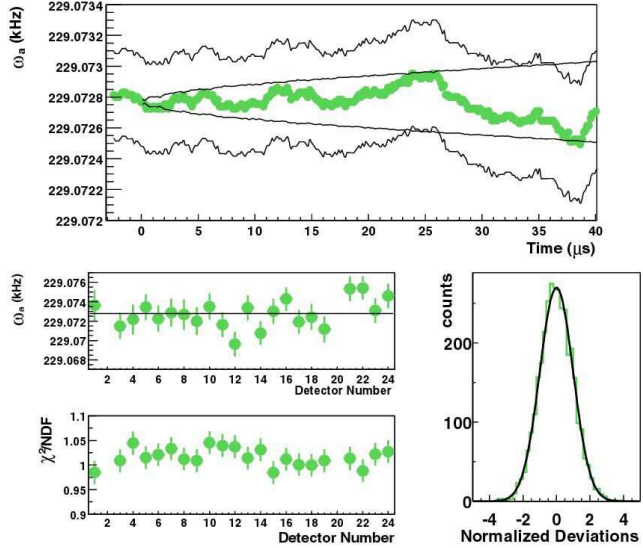}}
\caption{Internal consistency checks for muon precession frequency.
\label{anom}}
\end{center}
\end{figure}

Results from Experiment E821 at Brookhaven National Laboratory were
reported by Onderwater.~\cite{gerco} 
To measure the anomalous magnetic moment of the muon, polarized muons
from $\pi$ decay precess in a magnetic field produced by a storage
ring with a 1.45 T superconducting magnet uniform to 10 ppm. The
magnetic field is measured to 1 ppm by an NMR probe on a trolley. The
precession frequency of the NMR protons is $\omega_p/2\pi = 61,791,256
\pm 25$ Hz. The precession frequency of the muons is measured to be
$\omega_a/2\pi = 229,072.8 \pm 0.3$ Hz. There are several internal
consistency checks, shown in Fig.~\ref{anom}.

The result is  $a_\mu^{\rm exp} - a_\mu^{\rm theory} =$ ($43 \pm 16$)
$\times 10^{-10}$, a $2.6 \sigma$ deviation. The total error is
statistics dominated, with a systematic error of 0.25 ppm. Data
equivalent to four times that presented here have been recorded but
not yet analyzed, and the goal is to improve the final
error to 0.3 ppm. The largest contribution to the theoretical error is
the hadronic contribution,~\cite{hocker} which is being redone to
include the BES
$R$ measurement and new $\tau$ results from ALEPH. The deviation from
the Standard Model expectation is consistent with contributions from
loops due to supersymmetric particles.~\cite{czarn} It is hoped to have
a new result by the end of the year.

\section{Conclusions and Future Prospects}

The evidence for the Standard Model Higgs boson from LEP has a
significance of only $2.9 \sigma$. Unfortunately a LEP run for 2001
was not approved, so we will have to wait until at least $\sim$ 2007
to find out from the LHC experiments, or possibly the Fermilab
Tevatron, whether there is really anything there. The significance of
the deviation of the muon anomalous magnetic moment from the Standard
Model prediction is even less, $2.6 \sigma$; however, in this case we
should have more information soon from the analysis of the rest of the
E821 data from 2000. If both of these possibilities are verified, then
the evidence for supersymmetry will be rather strong. If the 115 GeV
Higgs boson is the lightest supersymmetric Higgs, we should be able to
find the heavier Higgs bosons or supersymmetric particles at the
LHC. However, there are cases in which the observation of the heavier
Higgses at the LHC or at a future $e^+ e^-$ linear will be difficult.

The evidence for neutrino oscillations seems to be strong; more
measurements will be reported in the near future. An exciting program
of neutrino physics awaits us, as we measure the masses and mixing
angles, and possibly even $CP$ violation.

\section*{Acknowledgments}
The author acknowledges support from the U.S. Department of Energy
grant DEFG0291ER40661. The author thanks Professor J. Tr\^an Thanh V\^an
and the conference organizers for a stimulating and enjoyable conference.


\begin{thebibliography}{99}
\bibitem{gen}
References are to talks presented at this conference. Further
references may be found in the individual contributions presented in
this volume.
\end{thebibliography}
\end{document}